# Adaptive Beaconing Approaches for Vehicular ad hoc Networks: A Survey

Syed Adeel Ali Shah, Ejaz Ahmed, Feng Xia, Ahmad Karim, Muhammad Shiraz, Rafidah MD Noor

*Abstract*—**Vehicular communication requires vehicles to self-organize through the exchange of periodic beacons. Recent analysis on beaconing indicates that the standards for beaconing restrict the desired performance of vehicular applications. This situation can be attributed to the quality of the available transmission medium, persistent change in the traffic situation and the inability of standards to cope with application requirements. To this end, this paper is motivated by the classifications and capability evaluations of existing adaptive beaconing approaches. To begin with, we explore the anatomy and the performance requirements of beaconing. Then, the beaconing design is analyzed to introduce a design-based beaconing taxonomy. A survey of the state-of-the-art is conducted with an emphasis on the salient features of the beaconing approaches. We also evaluate the capabilities of beaconing approaches using several key parameters. A comparison among beaconing approaches is presented, which is based on the architectural and implementation characteristics. The paper concludes by discussing open challenges in the field.**

*Index Terms*—**Vehicular ad hoc Networks, Adaptive Beaconing, Networking, Protocols, Intelligent Transportation Systems, Vehicle Safety Communications**

## I. INTRODUCTION

**T**HE past decade has seen the rise of a wireless communication technology that promises to improve the traveling experience on roads. Vehicles and road-side-units (RSUs) with sensing capabilities are now able to collect information about the road and traffic situations with exceptional detail and ubiquity. As the motor industry aims to equip vehicles with the Dedicated Short Range Communication (DSRC) technology, we may soon experience services of a sustainable Intelligent Transportation Systems (ITS) with an aim to make traveling safer, comfortable and environmentally friendly.

Large-scale ITS has a substantial possibility of deployment in the near future due to the serious intent shown by the research community to provide safety while traveling. A high-level perspective of safety in vehicular networks is the early driver's reaction to a potentially hazardous safety situation. A study showed that approximately 60 percent of accidents could be avoided provided that drivers receive warnings earlier by a fraction as low as half a second [1]. It follows that, awareness about the surrounding traffic situation is a crucial requirement for effective ITS.

Syed Adeel Ali Shah, Ejaz Ahmad, Rafidah MD Noor, Ahmad Karim are with Department of Computer System and Technology, University of Malaya, Malaysia, email:adeelbanuri@siswa.um.edu.my

Feng Xia is with School of Software, Dalian University of Technology (DUT), China, email:f.xia@ieee.org

Muhammad Shiraz is with Department of Computer Science, Federal Urdu University of Arts, Science and Technology Islamabad Pakistan, email: muh_shiraz.dr@yahoo.com

Usually safety applications demand vehicles to self-organize in order to maintain accurate awareness about the traffic situation. Therefore, vehicles use a high message frequency to periodically broadcast their status (speed and direction) using beacons within the 1-hop transmission range [2].

However, recent analysis on beacon transmission indicates that existing standards in beaconing restrict the performance of vehicular applications [3], [4]. Essentially, for constant message frequency, the limited wireless bandwidth causes loss and erroneous beacon reception under high-density networks. It is also worth mentioning that the standards for periodic beaconing are designed to be inline with the spontaneous mobile ad hoc communication requirements, which do not necessarily conform to the communication requirements of vehicular safety applications. That is, during high message frequency transmissions by safety applications, channel saturation and loss of messages cannot be addressed by the beaconing standards alone [5]. This situation calls for adaptive beaconing approaches that can efficiently utilize the wireless channel and provide reliable vehicular communications.

Despite the vast number of proposals, only a few surveys exist on beaconing approaches [6], [7], [8]. The study in [6] classified adaptive beaconing as means for controlling congestion and improving neighbor awareness by surveying a few beaconing approaches. Another study in [7], classified safety and non-safety applications to discuss inter-vehicle communication protocols, which also included multicast and broadcast. A recent survey also summarized salient features of beaconing approaches [8] along with some simulation results. In contrast, this study aims to provide a comprehensive survey of the developments in the area of adaptive beaconing from its initiation to the most recent proposals. Moreover, the survey is designed to be comprehensible for the readers even outside the specialty of the topic. Therefore, we have used a qualitative approach to conducting this survey, which provides discussions on the important concepts without putting complicated results into the context. The key contributions are listed as follows.

### A. Contributions

- The paper describes anatomy of beaconing through a schematic layered illustration and multi-channel communication perspective.
- We list the key performance requirements of beaconing and elaborate the beaconing design with respect to the information required by the beaconing approaches.
- With the aim of classifying beaconing approaches, the paper introduces a design-based taxonomy.

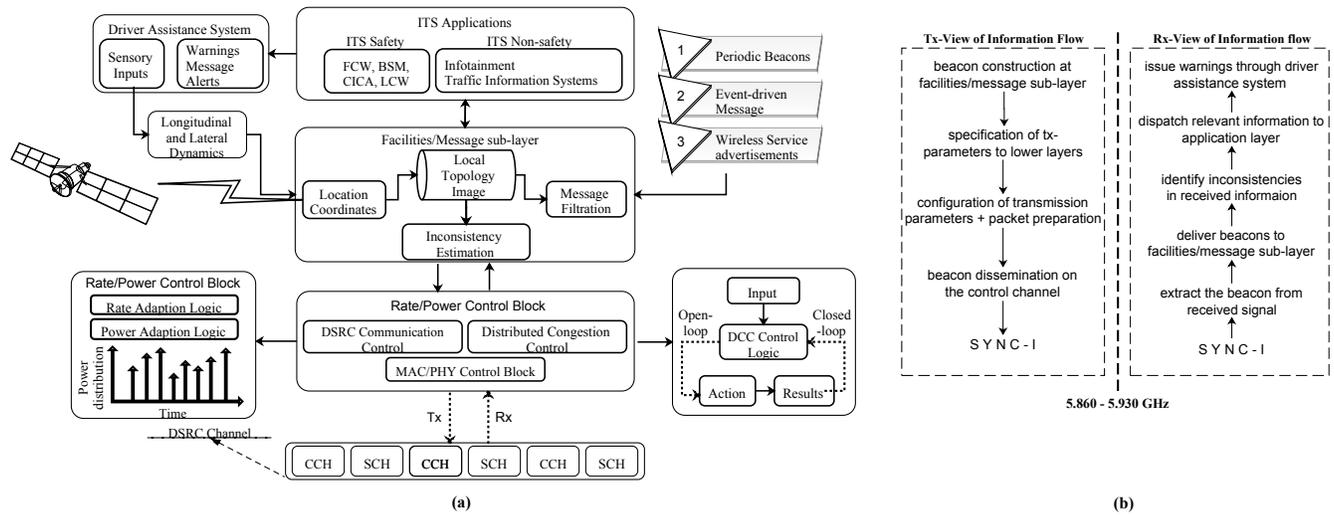

Fig. 1: (a) Schematic layered Illustration of Beaconing in Vehicular ad hoc Networks, (b) Transmitter and Receiver perspective of information flow

- A survey on the salient features of beaconing approaches is given to highlight the key observations about each category of the beaconing approach.
- We qualitatively evaluate the capabilities of beaconing approaches using important parameters.
- The paper explores the architectural characteristics to further classify and compare the beaconing approaches.
- Finally, we present some open challenges.

These contributions are given in separate sections from II to VIII; the conclusion is given in Section IX.

## II. ANATOMY OF BEACONING

Here, we describe the layered illustration and multi-channel communication perspective of 802.11p for beaconing [9].

### A. Schematic Layered Illustration of Beaconing

The European Telecommunications Standards Institute (ETSI) [10] and IEEE Wireless Access for Vehicular Environments (WAVE) [11], [12] have conceived the necessary layered architecture for vehicular application communication. In Fig. 1(a), we show a schematic illustration of this layered architecture, which does not necessarily correspond to the actual standard; rather, it highlights only the noteworthy aspects involved in beaconing. The flow of beacons is also illustrated with respect to the transmitter and the receiver in Fig. 1(b).

The beacons in VANETs provide the actual services for the safety and non-safety applications. However, upon close examination, it can be observed that the scope of the required information for vehicular applications is limited to sensor inputs, vehicular speeds and longitudinal and lateral dynamics, to name but a few. Accordingly, there exists an additional facilities layer/Message sub-layer. The role of this layer is to maintain a local topology image encompassing neighbor vehicles and to communicate with the application layer, which in turn informs the driver assistance system to generate warnings and message alerts. Two types of messages are used for this purpose: 1) periodic beacons for neighbor localization, and 2) event-driven messages to inform neighbors about a potential hazard. It should be noticed that beacon transmission is broadcast and vehicles may receive messages not intended for them. Therefore, this layer is also responsible for dropping irrelevant messages through message filtration. The inconsistency estimator is critical in identifying variations in the desired accuracy of the local topology image and in the longitudinal and lateral dynamics of the vehicle itself. The other type of messages is the Wireless Service Advertisements (WSAs), which are used to advertise non-safety application services by the service providers.

Regardless of the type of message, the efficient periodic beacon dissemination is governed by the choice of certain parameters, such as message frequency and transmission power along with several MAC/PHY layer parameters. Indeed, the choice for adapting these parameters can be based upon the traffic context and application-specific context. As such, the DSRC communication control block specifies how message frequencies and transmission powers are adapted, while the MAC/PHY control block allows the adaptation of contention windows, and data-rates to name but a few. Also, in order to avoid channel saturation in the wireless medium, the distributed congestion control block specifies congestion control strategies based on the open-loop and close-loop control theory approaches. After fine-tuning of the transmission parameters at the DSRC control block, the beacons are then disseminated on the shared wireless medium. Next, we examine the details involved in multi-channel beacon transmissions.

### B. Multi-channel Communications

To avoid the complexities of channel association and to provide spontaneous wireless access, 802.11p is specified as a multi-channel access mechanism for VANETs. It is based on 802.11x Wireless Local Area Network (WLAN) standard. Unlike WLAN, the 802.11p distributes its services across two types of channels [13]: 1) control channel (CCH) and 2) service channel (SCH). The periodic beacons are transmitted

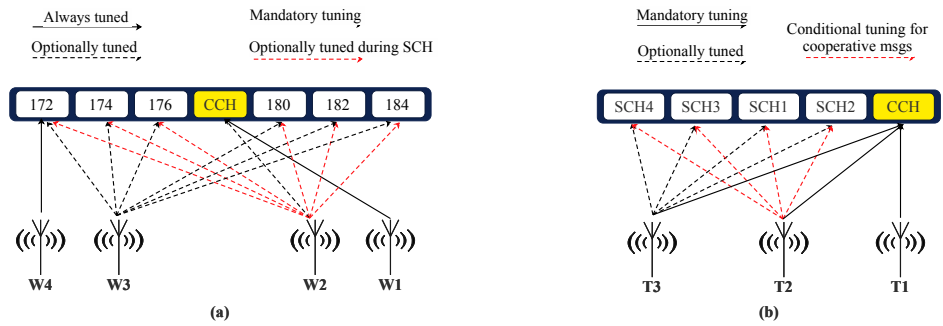

Fig. 2: Single and dual radio configurations for microscopic dissemination: (a) IEEE WAVE radio configuration options (b) ETSI radio configuration options

with a high frequency on the CCH. In addition, providers also advertise the non-safety applications on the CCH. In response to the non-safety advertisements, users can switch to a SCH to access a service. It implies that vehicles are bound to tune their radios to a desired channel for a particular service.

The Fig. 2 illustrates radio configurations for channel switching using single and dual radios [14]. The IEEE WAVE has one CCH and six SCHs with four different configurations as shown in Fig. 2(a). The W1 configuration requires a radio to permanently tune to the CCH. The W2 configuration requires the radio to switch between the CCH and SCH. In W3, the radio can switch between the available SCH but cannot access the CCH. In W4, the radio is permanently tuned to the CCH and cannot access any SCH. It should be noted that W3 and W4 configurations require dual radio setup to access both safety and non-safety services. A single radio tuned in W4 requires a second radio in W2 configuration. A radio tuned in W3 can have another radio which is tuned in configuration W1 or W2. Fig. 2(b) shows the radio configurations in the ETSI standard for one CCH and four SCHs. The T1 configuration permanently tunes the radio to the CCH. The configurations in T2 and T3 are used to access any SCH with mandatory tuning to the CCH. During congestion, the ETSI also allows service advertisements on the SCH1.

To provide service differentiation, these standards allow classification of messages by using Enhanced Distributed Coordinated Access (EDCA) of 802.11e [15]. According to EDCA, contention window sizes are assigned to different traffic categories. High-priority data gets lower contention windows and vice verse. In addition, the default data rate for beaconing is defined as 6 Mbps [16].

## III. Performance requirements and Beaconing Design

This section discusses the desired performance requirements and the subsequent design of adaptive beaconing approaches.

### A. Performances Requirements

The following performance requirements are desirable for beaconing approaches: 1) reduced beaconing load, 2) application diversity, 3) fairness in beacon transmission, and 4) reliable beacon delivery.

*1) Reduced Beaconing Load:* Beaconing load on the control channel indicates the state of channel occupancy. During the state of high occupancy, otherwise known as congestion, the vehicular applications scale poorly. For instance, beaconing load for dense networks can affect application performance due to higher latency in acquiring neighbor awareness. To manage the state of channel occupancy, beaconing approaches must adapt according to the vehicular density, channel states and Bit Error Rates (BERs) etc. Possible mechanisms to reduce load includes 1) the use of position prediction algorithms, and 2) reduction in transmission power. The former avoids unnecessary beacons by predicting the vehicle positions based on their previously received positions while the latter restricts the transmission range to minimize beaconing load. It should be noted that it is undesirable to use feedbacks for beacons which have marginal temporal validity.

*2) Application Diversity:* Vehicular networks have the capability to host safety applications [17], [18], [19] as well as non-safety applications [20]. These applications have diverse requirements for beacon dissemination and awareness levels. For instance, beacon dissemination for safety applications are delay-sensitive, while non-safety applications [21], [22], [23], [24], [25] can tolerate certain level of delay. Therefore, application diversity of a beaconing approach indicates the ability to satisfy different application requirements.

*3) Fairness in Beacon Transmission:* The consistency in mobility and communication patterns are pertinent to vehicular networks. Consequently, vehicles have different views about their respective topologies and the state of channel occupancy. Under these conditions, for a beaconing approach to be fair, it must perceive the views of neighbors before adapting a transmission behavior. In other words, the adaptive transmissions from one vehicle should not affect transmissions from its neighbors. As an example, consider a vehicle experiencing low channel occupancy, which subsequently increase its message frequency. However, beaconing approach with fairness criterion demands that frequency increase should not cause a state of congestion in neighbors and force them to use a lower message frequency. Similarly, an increase in transmission power is known to decrease the message reception probability of nearby neighbors [26]. Therefore, power adaptation must be fair across vehicles within a transmission range. Implementing fairness requires information sharing among vehicles about their views of the topology and network states.

TABLE I: Impact of gathered information from different information sources

| Information category | Beaconing load | Application diversity | Fairness | Reliability |
|---|---|---|---|---|
| App. Requirement | – | – | Direct | Direct |
| Network state | Direct | – | Direct | Indirect |
| Traffic scenario | Direct | – | – | – |
| Mixed | | Combination of above | | |

*4) Reliable Beacon delivery:* A critical objective of adaptive beaconing for safety applications is the reliable delivery of beacons in a timely manner. Note that, the beaconing approaches do not use acknowledgments to indicate beacon reception. Therefore, reliability must be provided through different adaptive approaches such as message prioritization, increasing message frequency and transmission power etc. Providing reliable delivery of safety-critical data may violate fairness criterion at some vehicles, therefore, an appropriate choice for measuring reliability is to use application-centric metrics such as Time-Window Reliability (T-WR) and drivers reaction time to name but a few.

### B. Design of Adaptive Beaconing Approaches

Two aspects govern the design of beaconing approaches: 1) the information used as input by the adaptive beaconing approach, and 2) the subsequent choice of the control mechanism for beacon transmission. Here, we restrict our discussion to the concepts and categories of the information used as input for the adaptive beaconing. Table I lists the impact of the type of information on the system performance requirements.

*1) Application Requirements:* Application requirements indicate the transmission requirements of beacons for correct operation of vehicular applications. Based on these requirements, the MAC layer can adapt the beacon transmission accordingly. For instance, in ETSI, the facilities layer can specify a strict temporal requirement for the MAC to transmit beacons with prioritized access using 802.11e. The application requirement has a direct impact on the fairness and reliability. Reliability results in improved reaction time for the drivers in a hazardous road condition e.g. through a prioritized delivery of a collision avoidance message.

*2) Network State:* Network state specifies the locally computed performance metrics of the wireless channels. It includes Channel Busy Ratios (CBR), bit error rates (BER) and interference levels, to name but a few. Most of the approaches depend on these metrics to cope with the scarce channel resources. It follows that using network state has a direct impact on the beaconing load and fairness. With the capability of sharing the network states with neighbors, the secondary impact can be specified in terms of improved reliability.

*3) Traffic Scenario:* Traffic situation corresponds to the one of the defined safety scenarios in [17]. To detect a traffic scenario, a vehicle constantly monitors the status of neighbors and road conditions [27]. Specifically, the vehicle yaw rates, intersection crossings, overtaking maneuvers and roads with merging locations etc. Variations in a traffic situation directly impact the beaconing load required for a certain level of awareness. For instance, a high-speed vehicle approaching an intersection needs a higher message frequency and hence a high beaconing load [28].

*4) Mixed:* The discussed information categories can be used in different combinations for a multi-objective beaconing approach. As an example, reduced beaconing load and reliability requires network state as well as application information in order to maintain CCH saturation and to provide timely delivery of beacons [27].

## IV. TAXONOMY

The design of beaconing approaches can be classified into 1) Beaconing category, 2) Information dependency and 3) Objective function as shown in Fig. 3. Each design element is discussed in the following.

### A. Beaconing Category

Beaconing category classifies the response by a beaconing approach to the acquired information. The four beaconing categories include: 1) Message frequency control (MFC), 2) transmit power control (TPC), 3) Miscellaneous and 4) hybrid.

With message frequency control, vehicles adapt the frequency for beacon transmission. The aim is to improve CCH utilization for a variety of objective functions. For instance, lower frequency reduces CCH load and helps achieve a higher probability of beacon reception. Similarly, a higher frequency can be used to guarantee message delivery in an intersection collision warning system [28].

Transmit power control is primarily used as a topology control mechanism. The TPC approaches share similar objective functions with the MFC. Additionally, efficient TPC increases the throughput, coverage area of the transmitter [29] and the reception probability for specific regions [30].

More recently, researchers have started focusing on the multi-channel switching aspects of 802.11p for adaptive beaconing. We refer to these aspects as Miscellaneous, which includes an adaptation of contention window size, physical data rates and de-synchronization of transmission intervals etc.

Hybrid control specifies the combination of MFC, TPC, and miscellaneous approaches. The aim is to exploit the strengths offered by different approaches. For instance, given the fundamental bounds of wireless networks, transmission rate and power must be adapted to efficiently utilize shared channels. In VANETs, this notion is relevant and may be utilized according to the context and desired objectives. Such as, transmit rate can be adapted according to the traffic context while power can be adjusted for higher information-penetration.

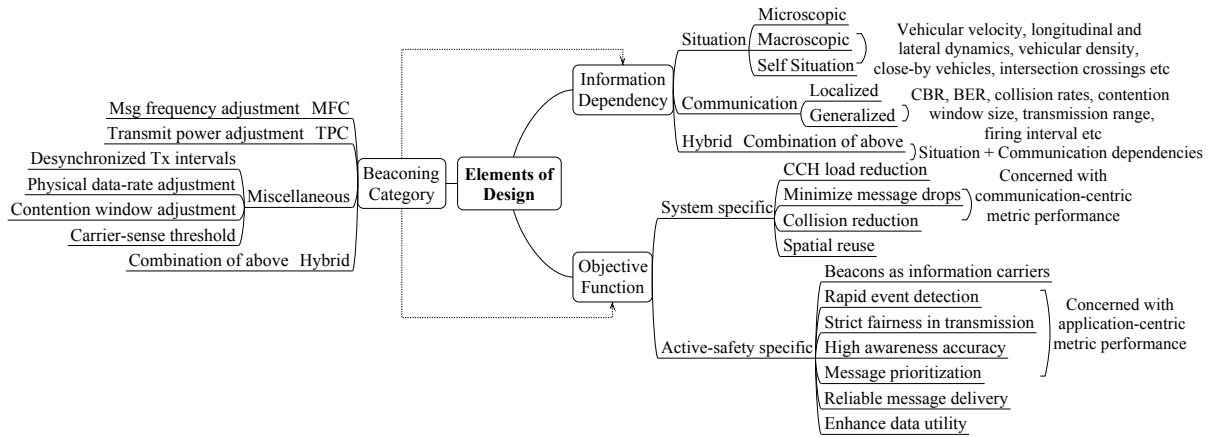

Fig. 3: Design-based Taxonomy of adaptive beaconing approaches

## B. Information Dependency

The beaconing approaches can also be classified by examining the information required by a beaconing approach for decision making. This information represents 1) situation, 2) communication feature and 3) hybrid.

The first category indicates the traffic situation. That is, at a particular time, the vehicular network represents a unique traffic situation. It further implies that the traffic information is diverse and can be classified as microscopic, macroscopic and self-situation. Microscopic information refers to the exact traffic scenarios as specified by ETSI [17]. On the other hand, macroscopic information describes an overall traffic situation such as vehicular density i.e. high or low. Finally, self-situation represents the variations in the movement of a vehicle itself.

Communication feature specifies the transmission characteristics. For example, CBR, clear channel assessment reports, interference levels, and packet collision etc. It follows that this information can be differentiated as 1) localized and, 2) generalized. The former specifies local computations of transmission characteristics while the latter specifies local as well as acquired statistics from the neighbors.

The hybrid category is more diverse and uses a combination of traffic-related information and communication information.

## C. Objective Functions

The objectives of beaconing can be system-specific or application-specific. System-specific objectives include performance improvement in the quality of transmission. For instance, reducing the channel load on CCH, managing packet drops, and controlling collisions during transmission. System-specific objectives require localized or generalized communication information.

On the other hand, application-specific objectives aim at enhancing safety application performance. For instance, rapid event detection, fairness in channel access, a higher degree of awareness, prioritization of critical event-driven messages and message delivery within strict time constraints.

## V. Survey of Adaptive Beaconing Approaches

Here, we survey the most important beaconing approaches in existing literature. At the end of this section, in Table II, we only summarize the key observations about each surveyed beaconing category due to space limitations.

## A. Message Frequency Control Approaches

With message frequency control (MFC), vehicles adapt the frequency for beacon dissemination. In this section, we survey the most noteworthy MFC proposals in literature.

*1) MFC based on traffic situation:* The most common criterion for adapting message frequency is the surrounding traffic situation. As such, the MFC approach in [27], studied the vehicular density and speed for message frequency adaption. The authors proposed to adapt message frequency based on a vehicle's own movement or based on the surrounding situation. A vehicle's own movement includes speed and yaw rates along with special vehicles that need prioritized access to the road lane. Any of these conditions require a higher message frequency. By contrast, congestion is deemed more significant than awareness in high-density networks because of the high probability of beacon collisions. Therefore, the study proposed to reduce the frequency based on vehicular density to reduce congestion and maintain an acceptable level of awareness.

*2) MFC based on position prediction:* The MFC approaches in [31] and [32] proposed to use a vehicle position prediction model for adapting the message frequency. The basic motivation is to reduce congestion by minimizing unnecessary beacon transmissions.

The MFC in [31] models position prediction by exploiting the information in the received beacon. That is, after receiving a beacon from the neighbor, the receiving vehicle starts predicting the neighbor's position for a specified time. In the meantime, beacons are sent to the predicted neighbor position. The position is updated upon the reception of a new beacon from the same neighbor.

Similarly, the position prediction in [32], is based on the Kalman filter for distributed position estimation logic. It means that after advertising a beacon, the vehicle calculates its own position in the near future that replicates the position calculation process at the neighbor vehicles. Subsequently, using the errors in position, the vehicles adjust their message frequency.

*3) MFC based on fairness:* Fairness in adapting message frequency is critical and requires cooperation among vehicles. The approaches proposed in [33] and [3] provide fairness with respect to the periodic beacons and the event-driven messages.

The approach in [3], and [4] use high channel occupancy as an indicator for high congestion. Therefore, during congestion, a vehicle informs the neighbors about its state. Upon reception of this message, all vehicles cooperate by blocking the transmission of periodic beacons to allow the transmission of potential event-driven messages. Moreover, in response to the message, vehicles adapt the message frequency by using the concept of additive increase and multiplicative decrease. This approach implies that the message frequency is initially increased by one message per second and reduced by half if congestion occurs.

Similarly, periodically updated load-sensitive adaptive rate control (PULSAR) [33] adapts message frequency by considering the vehicles that cause congestion within the carrier sense range. For a specified time interval, a vehicle monitors the CBR and listens to the CBR advertised by the neighbors. Thus, the adapted message frequency maintains an acceptable CBR level to provide highly probable transmission of event-driven messages. Moreover, the feedback helps in identifying vehicles that contribute more to the congestion. As a result, the transmission rate of such vehicles is reduced.

*4) MFC for overtaking assistance:* An important consideration in adaptive MFC is the performance of safety applications. The approach in [34] defines two types of vehicles: leading and regular vehicles. Unlike regular vehicles, the leading vehicle has no neighbors in front. Therefore, to enable safe overtaking maneuver by regular vehicles, the leading vehicle monitors, and reports the presence of oncoming traffic. For reliable reporting, this information is transmitted by using event-driven messages at a higher message frequency than the normal vehicles.

### B. MFC for non-safety applications

Non-safety applications may also benefit from message frequency adaptation. The authors in [35] proposed to use beacons to transmit Traffic Information Systems (TIS) data [36]. The main objective is to provide high event-penetration ratio without using flooding. To reduce the channel load, message frequency is adapted based on two pieces of information: 1) the distance of the vehicle that generates the TIS data to the event and 2) the age of the disseminated message that specifies the information freshness. It also uses communication-driven information that includes 1) the number of collisions, 2) the signal-to-noise ratio, and 3) the number of received beacons. To ensure that rate adaption is inclined toward congested channels and collisions, communication-driven parameters are weighted more than the distance to the event and message age parameters. This situation implies that the highest transmission rate is selected if the TIS data is fresh and the channel usage is minimal. Otherwise, the transmission rate is optimized to meet the dissemination requirement of the data while keeping the channel load at a minimum.

The beaconing approach in [37] controls the frequency of beacons for efficient bandwidth utilization. In addition, the approach introduces a fair data selection mechanism such that the most significant messages receive high priority for transmission. This condition is achieved by identifying vehicle interests in the data and then distributing that interest among the neighbors. Moreover, to efficiently utilize the channel, message frequency is controlled by considering parameters such as data age, distance to the destination/roadside unit, history of message reception, and vehicular interest in data.

### C. Transmit Power Control Approaches

Transmit Power Control (TPC) defines variation in the transmission power for beacon dissemination. In this section, we survey the TPC approaches proposed in [38], [39], [40], [41] for VANETs.

*1) TPC based on fairness:* For constant message frequency, the authors in [38] defined the congested region as the one with the maximum number of interfering interference ranges. The motivation of the proposed TPC is to provide strict fairness in the transmit power through cooperation. The approach works in two phases. In the first phase, a vehicle collects information about the power levels of the neighbors. The vehicle then calculates a power value, which is the maximum common value among the received power values. In the second phase, the calculated power value is advertised. This step ensures that the locally computed power level does not violate the congestion requirement of the neighbors. Finally, upon reception of the calculated power values, the vehicle selects the minimum power level to transmit.

*2) TPC with random power level:* In highways, vehicles have a tendency to form clusters because of minimum relative speed variations. Under this scenario, beacon collisions are recurring. As such, the approach in [40] proposes random transmit powers to reduce recurring collisions to increase neighbor awareness. To ensure fairness in power selection, all vehicles randomly select transmit powers by using a common mean and variance. The mean power level enables the vehicles to maintain a higher awareness of close-by neighbors. Furthermore, it reduces the overall congestion by transmitting less at longer distances.

*3) TPC for spatial reuse:* Power adaption can be used to optimize the spatial reuse and to provide transmission over long distances. The TPC in [39] provides spatial reuse by reducing the transmit power for beaconing. Initially, a beacon is transmitted by using a low power level. It is followed by the retransmission phase, which also provides a simple form of information aggregation, that is, along with the received beacon, the relay also transmits self-information. To retain information freshness and to avoid delays in the multihop transmission, the retransmission is scheduled on selected relays.

*4) TPC based on feedback:* The authors in [41] proposed an application-based power control using feedback. The motivation was to transmit safety messages with enough power level to cover the desired range with no excessive coverage. The initial transmit power for beacons is assigned with respect to the coverage required by an application. Each vehicle then maintains a speaker list, which contains neighbors whose power level exceeds the desired coverage. A feedback message is used to notify the speaker list about their high transmit power levels. Therefore, neighbors whose addresses are included in the feedback reduce their transmit power.

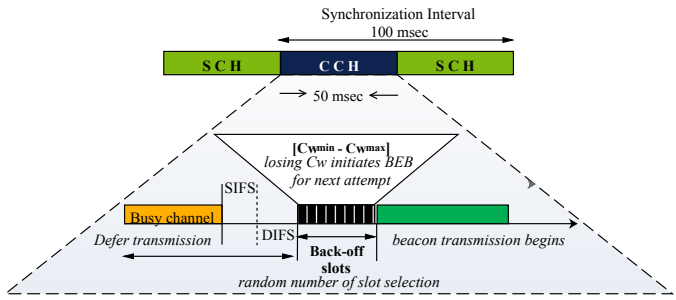

Fig. 4: Illustration of the access mechanism using contentions and back-offs

### D. Miscellaneous Approaches

Miscellaneous approaches include a variety of parameters for the adaptation of beaconing, which specify intricate details of the MAC layer as shown in Fig. 4. In the following, we first describe these parameters before proceeding with the survey.

The distributed inter-frame space (DIFS) is a time interval for a vehicle to wait before transmission. Note that, the Short Inter-frame Space (SIFS) precedes the DIFS and it represents the collective time required to process a received frame and a subsequent response frame. Before transmission, if the medium is sensed idle for the duration of DIFS, the vehicle starts transmitting. Otherwise, the vehicle enters a contention period by choosing a back-off. Upon the expiry of the contention period, the vehicle can start transmitting if other vehicles have selected higher back-off values. If a vehicle still finds the channel busy then as part of Binary Exponential Back-off (BEB), the contention window size is increased for the next transmission attempt. The rate of beacon transmission is specified through the data rate to effectively utilize link throughput. Another parameter is the guard interval, which is a 4 msec timer between the CCH and SCH interval. During this interval, the channel is advertised as busy to restrict transmission. In the following, we survey some of the important beaconing approaches in this category.

*1) De-synchronized beacon transmissions:* A stochastic model for non-deterministic channel switching and its effect are introduced in [42] with an aim of reducing collisions. The approach monitors the beacon firing intervals used by the neighbors. Then, a vehicle selects a distinct firing interval for beacon transmission. In addition, to minimize the probability of converging at the same firing interval, it uses random back-off before transmitting at the selected firing interval. It also has a provision for a network having different periodic transmission intervals, that is, vehicles can report their transmission time periods to each other. In this way, any vehicle that receives a beacon sets its own time period, which is multiples of the received time period. This approach allows adaptability for vehicles having shorter transmission intervals with those having longer transmission intervals.

More recently in [43], the authors proposed an application-based mechanism to reduce MAC layer collisions. At the application layer, they proposed the intuition of replicating the CCH interval with consecutive 1 ms epochs. Vehicles keep track of the epochs used by the neighbors in the previous time interval. To avoid possible collisions, a vehicle uses an underutilized epoch. In situations where all the epochs are utilized, a random epoch is selected for beacon transmission.

*2) Adapting physical data rate:* In Vanets, the link quality changes due to fading and mobility. Therefore, an accurate estimate of bit rate adaptation is required to effectively utilize the link throughput with respect to the link quality.

To gain maximum utilization of the link, the approach in [44] proposes an estimation of the data rate for transmission. It exploits the local information for rate adaption without constant probing of the link. As a result, this approach induces minimum delays in rate adaption. With the ability to exploit local information, faulty data rate selection at the beginning of the bursty traffic could be avoided. Data rate estimation is based on two types of functions. The first uses the current context, that is, the local topology, current data rate, and the packet size to estimate packet error. The second function uses previous statistics on the bit rates as exponentially weighted moving averages. The estimation of packet error in the first function for beacon dissemination is based on an empirical model, which uses multivariate linear regression on the measurements obtained from real test beds. For high-speed vehicular networks, data rate estimation using the context is given higher preference over the estimates of previous statistics on the bit rate.

The objective of beaconing approach in [45] is to avoid synchronized collisions at the beginning of the channel interval and to minimize packet drops before the end of the control channel interval. The solution is based on two observations: 1) the beacons can collide due to same back-off selection by different vehicles at the start of a channel interval and 2) a packet may be dropped if the required transmission time for the packet exceeds the available transmission time of the channel interval. To handle the first observation, a longer contention window is proposed, which brings diversity in the slot selection and higher variation in the selection of waiting times before transmission. Nevertheless, this approach reduces the interval for the transmission of beacons and a higher probability of packet drops before transmission. Therefore, the approach introduces a higher data rate. Before transmitting at higher data rates, the delivery probability is calculated for an assured reception.

*3) Adapting contention window:* In the 802.11p standard, beacons are transmitted with the access category (AC) IV. Due to the short temporal validity of beacons the minimum contention window size of AC-VI is kept small. Recent studies have shown that the minimum window size of AC-VI is the main cause of beacon collisions. Here, we discuss recent solutions in handling collisions through contention window adaptation.

The approach in [46] avoids synchronous CCH collisions and addresses packet drops before transmission. The beaconing approach augments the exponential random back-off with slot utilization estimation. Slot utilization is based on the 1) current slot utilization, which is the average of busy slots to the number of available slots, and 2) previous values of slot utilization. After expiry of the back-off, the beacon transmission proceeds if the probability of successful transmission based on the available slot is sufficient. Otherwise, the beacon is dropped. Deferring beacon transmission due to increased contention window can compromise certain vehicles. Therefore, to enable fairness in packet drops across vehicles, a weighted probability of beacon transmission is calculated, which is based on the number of un-transmitted beacons and the number of transmission attempts for a beacon.

A related problem with the adaptation of contention window is the aggressive selection of back-off under normal conditions. In [47], the authors investigated the effect of contention window under various vehicular densities. Their contribution can be divided into two parts: 1) An analytical framework, which models the behavior of IEEE 802.11p MAC protocol, where the authors show that the broadcast nature of the safety messages affects the optimal value of the contention window. That is, a larger contention window is desired for high-density networks. Moreover, the contention window adaptation should aim to balance out the collisions and expired messages at the source. 2)A unique reverse back-off proposal, in which the initial contention window is set to a higher value. Then, based on the expired message, the window is reduced to half and vice versa for successful transmission.

*4) Beaconing based on carrier sense Threshold:* In multi-channel access, before transmission, the carrier is sensed to conclude a free or occupied channel. A high threshold suggests that the radio is less sensitive to transmission from the neighbors and vice versa.

R. K. Schmidt et al. [48], presented a stepwise clear channel assignment threshold adaptation for beaconing. The adaptation mechanism is based on the current waiting time of a beacon in a queue. That is, when a beacon arrives at t0, the default value is assigned to the clear channel assessment threshold. If the beacon stays in the queue after time t1, the threshold is incremented with an offset. After increasing the threshold, the clear channel assessment is carried out immediately. If the channel is found busy at time t2, the threshold is increased again. Finally, the procedure ends if the message is dropped or sent. Furthermore, the approach is capable of assigning priority to different types of messages. It also proposes to use a traffic-shaping mechanism by employing a mechanism similar to the token bucket scheme to regulate bursty traffic.

The authors in [49] proposed a receiver-initiated MAC protocol (RIMAC) for efficient spatial reuse. Unlike sender-oriented approaches, RIMAC allows the receiver to initiate transmission from the sender. The intuition follows from the fact that a sender cannot accurately sense the channel states of the receiver. Therefore, the effectiveness of both physical and virtual carrier sensing is employed in RIMAC. The receiver initiates the transmission by sending a short message that serves two purposes: 1) initiate transmission at the sender and 2) serve as a virtual carrier sense for vehicles besides the sender. To avoid a collision, physical carrier sense is used before transmitting the request to the sender. On the contrary, virtual carrier sensing allows the receiver to identify any existing request made by other vehicles. If detected, the transmission is delayed until the channel becomes free.

### E. Hybrid Control Approaches

Hybrid control approaches specify the combination of MFC, TPC and miscellaneous approaches. This section surveys hybrid approaches in [50], [51], [52], [53], and[28].

Vehicular movements can be represented as a tracking problem [54], which is used for the adaptation of message frequency and power in [51], [52]. The objective is to keep a free channel for high-priority data. Message frequency is regulated by using the error in the predicted neighbor position. As for power adaptation, vehicles monitor CBR. For a higher estimate of CBR, a vehicle assumes similar values for neighbors. This assumption helps in adapting the transmission power to reduce congestion on the channel.

A novel concept of beaconing as a service (BaaS) is proposed in [50]. BaaS uses vehicular distance to adapt message frequency and transmit power. It specifies the 100 m distance as critical for safety applications with a higher message frequency. A 2 Hz message frequency is used in excess of 100 m. Within the 100 m range, a collision partner is defined as a vehicle that has excessive longitudinal and lateral dynamics. Therefore, to avoid a possible collision, BaaS requests for a higher transmission rate from the collision partner. The request includes specifications of the desired transmission rate and the transmission duration. This approach uses a dual radio setup. Therefore, to handle adjacent channel interferences (ACI), the transmission power is reduced in one of the channels during parallel transmissions.

To satisfy the requirements of safety applications, a hybrid beaconing approach is introduced in [53]. In this approach, rate and power control regulates the CCH load to address the requirements of safety applications. Specifically, a lane changing scenario is considered for which the proposed approach detects an oncoming vehicle to avoid a potential collision. In this situation, the requirement of lane change application is reliable beacon delivery. To implement adaptive control, the beaconing approach uses the critical distance between vehicles for which a beacon must be shared to avoid a collision. This critical distance is considered between two types of vehicles: a) a vehicle initiating a lane change maneuver, and b) a vehicle representing potential collision during the lane change. To provide sufficient time for the driver to react, adaptive transmission emphasizes the reliable delivery of at least one warning message for vehicles entering the critical distance. Authors have proposed to use a higher transmission power with a low transmission rate. This concept is due to the fact that increasing the transmission rate reduces the transmission range in comparison to the increase in transmission power.

Another beaconing approach in [28] is designed for intersection crossings in urban scenarios. The main objective is to avoid vehicle collision at the intersection. The adaptation uses message frequency and transmit power. The algorithm becomes active when two vehicles from two different road segments reach a critical distance. Just like in [53], reliability is considered for a critical distance between vehicles in which they must receive one beacon to avoid a collision. Upon arriving at the critical distance, vehicles increase the transmit power, such that the probability of successful reception could be guaranteed for a given message frequency. The approach is more reliable than a similar evaluation conducted in [53].

TABLE II: Key observations about surveyed beaconing approaches

| Approach | Idea/Parameters for Adaption | Key observations |
|---|---|---|
| Message Frequency Control (MFC) | -Message frequency<br>-Situation prediction<br>-Fairness<br>-Safety application specific<br>-Non-safety application specific | -Message frequency adaption is disputed in context of stringent frequency requirements of safety applications<br>-Control mechanisms for position prediction lack timely prediction of potential hazardous situations<br>-Evaluation of tolerable extent up-to which frequency could be adapted for safety applications is a challenge |
| Transmission Power Control (TPC) | -Fairness in power allocation<br>-Random transmit powers<br>-Spatial reuse<br>-Exact transmission range | -A trade-off exists between fair power allocation and high beaconing load<br>-Power reduction requires consideration for propagation effects i.e. shadowing/fading<br>-Reduced transmit power brings synchronous collision close to the critical safety range specified by the application<br>-Unfair power allocation may cause low reception probabilities for vehicles at close range |
| Miscellaneous (Misc) | -De-synchronized transmissions<br>-Physical data-rate adaption<br>-Contention window adaption<br>-Carrier-sense thresholds based beaconing | -For higher PDR, data-rates beyond 6 Mbps is not beneficial due to sensitivity of modulation schemes to noise<br>-Increase in contention window increases the probability of beacons being dropped at the source<br>-Increase in contention window also helps in selection of different back-offs, hence reduced probability of collisions<br>-Being hardware-specific parameter, the carrier-sense threshold has very few proposals |
| Hybrid | -Combination of above | -Useful for highly specialized vehicular scenarios with hard QoS requirements<br>-Allows flexibility in choice of adaption parameters |

A transmit power and contention window adaptation is proposed in [55]. The idea of power adaption is based on an accurate estimation of vehicular density, whereas contention window adaptation is used in case of high-priority beacons. After estimating the vehicular density, the beaconing approach uses a static look-up table to select a power level, which is sufficient to cover the desired range. The number of collisions indicates the level of congestion on the channel. As a result, the contention window size is constantly adapted in direct proportions for all the access categories.

## VI. CAPABILITY EVALUATION

This section evaluates the capabilities of the beaconing approaches by using qualitative parameters as shown in Table III. The given parameter values are not absolute by any means. Instead, we use qualitative values to evaluate beaconing approaches. The aim is to qualify the evaluation parameters by describing the notion of the choice of a value for each parameter as discussed below.

### A. Beaconing Load

Beaconing load is significant in indicating the expected channel occupancy of a beaconing approach by evaluating their respective message frequency requirements. We measure beaconing load on a qualitative scale with values of low, application-dependent, and high. The beaconing approaches that restrict or defer beacon transmissions are most effective in reducing the beaconing load. The beacon transmission can be restricted based-on criteria such as by predicting the neighbor positions or low successful transmission probability of beacons. The application-dependent value implies that the beaconing approaches have no mechanisms to restrict or defer transmissions, rather the beaconing load is defined by the application requirements such as pre-crash sensing application that require 50 Hz message frequency. Finally, a high beaconing load is associated with: 1) beaconing approaches with multi-hop transmissions, 2) beacons that carry information as extra payload, and 3) beaconing approaches using feedbacks.

### B. Congestion Control Strategy

Congestion control strategy has a profound effect on the desired performance of vehicular applications. Therefore, the key objective of each beaconing approach is to minimize the state of channel occupancy. Beaconing approaches employ either an open-loop or a closed-loop approach to tackle high channel occupancy. The open-loop strategy is proactive, which requires an efficient design to stop congestion from occurring in the first place. As an example, adapting transmission power based on a predefined maximum beaconing load is a proactive congestion control strategy [38], which has a tendency to maintain channel occupancy at a certain level. As a result, unforeseen safety events can be transmitted with a higher probability of reception on a less congested channel. On the other hand, the closed-loop strategy is reactive, which allows the congestion to occur. Therefore, it requires feedback and continuous channel sensing mechanism to adapt transmission behavior. For instance, the approach in [3] uses channel busy time as an indicator for congestion and only then triggers the reduction in message transmission frequency. For spontaneous communication requirements of vehicular safety applications, an open-loop congestion control strategy is more desirable.

### C. Fairness

As discussed previously, the safety of each vehicle in a network is critical, therefore beaconing approach must ensure

TABLE III: Evaluation of adaptive beaconing approaches

| Category | Objective | Ref | S - NS | BL | CCS | Fairness | Reliability | DUD | CMDE |
|----------|-----------|-----|--------|-----|-----|----------|-------------|-----|------|
| MFC | traffic situation | [27] | ✓- ✗ | app. dep | y/CL | tx-fairness | probable | NA | NA |
| | position prediction | [31] | ✓- ✗ | low | y/OL | NA | best-effort | NA | NA |
| | position prediction | [32] | ✓- ✗ | low | y/OL | NA | best-effort | NA | NA |
| | fairness | [4] | ✓- ✗ | app. dep | y/CL | FCC | best-effort | NA | NA |
| | fairness | [33] | ✓- ✗ | high | y/CL | FCC | best-effort | NA | NA |
| | overtaking assistance | [34] | ✓- ✗ | app. dep | NA | tx-fairness | deterministic | NA | NA |
| | traffic penetration ratio | [35] | ✗- ✓ | high | y/CL | NA | best-effort | NA | NA |
| | efficient bandwidth utilization | [37] | ✗- ✓ | low | y/CL | NA | best-effort | yes | NA |
| TPC | fairness | [38] | ✓- ✗ | high | y/OL | tx-fairness | probable | NA | NA |
| | reducing recurring collisions | [40] | ✓- ✗ | app. dep | y/OL | NA | best-effort | NA | NA |
| | spatial reuse | [39] | ✓- ✗ | high | NA | NA | best-effort | NA | NA |
| | transmission coverage requirement | [41] | ✓- ✗ | high | NA | NA | best-effort | NA | NA |
| Misc | collision reduction | [42] | ✓- ✗ | app. dep | NA | NA | best-effort | NA | NA |
| | collision reduction | [43] | ✓- ✗ | app. dep | NA | NA | best-effort | NA | NA |
| | link throughput utilization | [44] | ✓- ✗ | app. dep | y/CL | NA | best-effort | NA | NA |
| | link throughput utilization | [45] | ✓- ✗ | app. dep | y/CL | NA | probable | NA | NA |
| | collision reduction | [46] | ✓- ✗ | app. dep | y/CL | NA | probable | NA | NA |
| | collision reduction | [47] | ✓- ✗ | app. dep | y/CL | NA | probable | NA | NA |
| | congestion reduction | [48] | ✓- ✗ | app. dep | y/CL | NA | best-effort | NA | NA |
| | spatial reuse + network capacity | [49] | ✓- ✗ | app. dep | No | NA | best-effort | NA | NA |
| Hybrid | tracking accuracy+congestion reduction | [52] | ✓- ✗ | app. dep | y/CL | NA | probable | NA | NA |
| | forward collision warning + ACI | [50] | ✓- ✗ | app. dep | NA | NA | deterministic | NA | NA |
| | reliable lane change warning system | [53] | ✓- ✗ | app. dep | NA | NA | deterministic | NA | NA |
| | reliable intersection collision warning | [28] | ✓- ✗ | app. dep | NA | NA | deterministic | NA | NA |
| | density estimation+msg prioritization | [55] | ✓- ✗ | app. dep | NA | tx-fairness | probable | NA | NA |

S-NS: safety non-safety, BL: beaconing load, CCS: congestion control strategy, DUD: data utility dissemination, CMDE: co-existing message dissemination evaluation, app. dep: application dependent, y: yes, CL: closed-loop, OL: open-loop, FCC: fair congestion control, tx: transmission, NA: not applicable that transmissions from one vehicle do not interrupt transmission from other vehicles. Providing fairness can be considered from two perspectives: a) fairness during transmission, and b) fairness in congestion control. The former approves fairness through a mutual agreement among vehicles to use certain adaptive behavior for general purpose periodic beacons. While the later, is applicable upon reception of congestion events from neighbors. Fairness in congestion control demands that the reduction in the frequency/power is relative to the vehicle's marginal contribution towards congestion. Note that, transmission of event-driven messages may violate fairness under certain conditions. For example, fairness criterion does not hold true for beaconing approaches that are designed to fulfill application-specific transmission requirements such as prioritized delivery and timely reception of messages (i.e. fairness is conditional). Under such conditions, the safety events are transmitted with the highest message frequency or transmission power, which may interrupt transmissions at certain vehicles within the transmission range.

### D. Reliability

In the absence of acknowledgments and because of the short temporal validity of beacons, reliability is treated as a critical performance measure for safety applications. We evaluate the reliability of beaconing approaches by using three scales: 1) deterministic, 2) probable, and 3) best-effort. The deterministic reliability is a guaranteed delivery of messages within strict time constraints, such as for lane change warnings approaches presented in [28], [50], [53] and [49]. Nevertheless, deterministic reliability has hard QoS requirements to fulfill. Therefore, such beaconing approaches cannot be generalized for a wide range of safety applications. Another most commonly used beaconing approach is to provide highly probable reliability by relying on the prioritization mechanism of 802.11p without strict time constraints. For example, contention periods for messages could be reduced by assigning a higher priority to beacons, such as in [38], [48], [47]. The rest of the approaches are considered best-effort, which supports only general purpose neighbor localization. These approaches are not suitable for safety applications.

### E. Data Utility

Data utility specifies the benefit of a received beacon for a vehicle. Enhancing data utility requires effective distribution of vehicles' interest in the data, which also specifies the conflict of interest among vehicles. As an example, in a two-way highway, vehicles in one direction may hold data relevant to vehicles in the other direction. Considering the available capacity for the exchange of only two beacons, the focus of a data utility-based approach is to select and forward messages that maximize the utility for receiving vehicles and not necessarily to improve the total utility of all vehicles [56]. The classification of vehicles based on their interests is crucial in saving bandwidth by transmitting data with the highest utility. The approaches in [56], [37] utilize data utility for the distribution of TIS.

### F. Co-existing Message Dissemination Evaluation

The notion of co-existing message dissemination evaluation originates from the conditional fairness property. As discussed previously, the fairness condition for periodic beacons is subject to a violation during the transmission of high priority event-driven messages. This concept applies particularly in situations where periodic beacons, event-driven messages, and wireless service advertisements co-exist in a network. Therefore, the idea of evaluating the effect of adapting transmission behavior for one type of periodic message on the other is subject to the empirical evaluation. To the best of our knowledge, [57] is the only study that explored co-existing periodic beacons and event-driven messages. In the study, the effects of varying priority levels and message frequency were evaluated. Its findings indicated that keeping periodic beacons at a lower priority enhances beacon up-to-dateness (metric for measuring awareness quality). By contrast, while event-driven messages transmitted at a higher frequency, the temporal effects, that is, low up-to-dateness can be monitored for the low-priority beacons. In light of this study, similar evaluations can be performed to analyze the periodic communication performance when parameters such as transmit power, contention windows, and physical data rates are adapted.

Based on the safety application requirements, beacon dissemination beyond a certain specified range is of least interest to the safety applications and is unreliable. By contrast, non-safety applications require periodic transmission of WSAs to a longer range for maximum advertisement coverage. In addition, the standards specify the use of local information for a provider to maintain an accurate view of SCH utilization in nearby providers. This context requires evaluations of different beaconing approaches on the performance of infotainment service providers and users alike. That is, significant insights could be gained by considering evaluation aspects, such as false channel selections by the users and suboptimal SCH utilization views by providers because of increased/decreased frequency, transmission power, contention window, and so on.

In order to evaluate beaconing approaches with respect to different messages that co-exit, a $benefit/cost$ approach could be used. This simple yet intuitive approach specifies the $benefit$ as a combination of metrics which are advantageous to different types of periodic messages. The tradeoff can then be represented using $cost$. For instance, the $benefit$ for safety application includes driver's reaction time, beacon up-to-dateness and per vehicle throughput etc. Similarly for TIS applications, penetration ratio and time can specify the benefit [35]. The $benefit$ for infotainment application may include WSA coverage and false channel selections to name a few. The $cost$ includes beacon collision rate, beacon drop probabilities, false channel selection rate and overhead etc.

## VII. COMPARISON

Based on the architectural and implementation aspects, this section presents a classification for beaconing approaches and conducts a comparison.

### A. Classification Approach

Upon close examination, several variations can be identified among beaconing approaches such as coordination mechanisms, resource utilization and implementation logic as shown in Fig. 5.

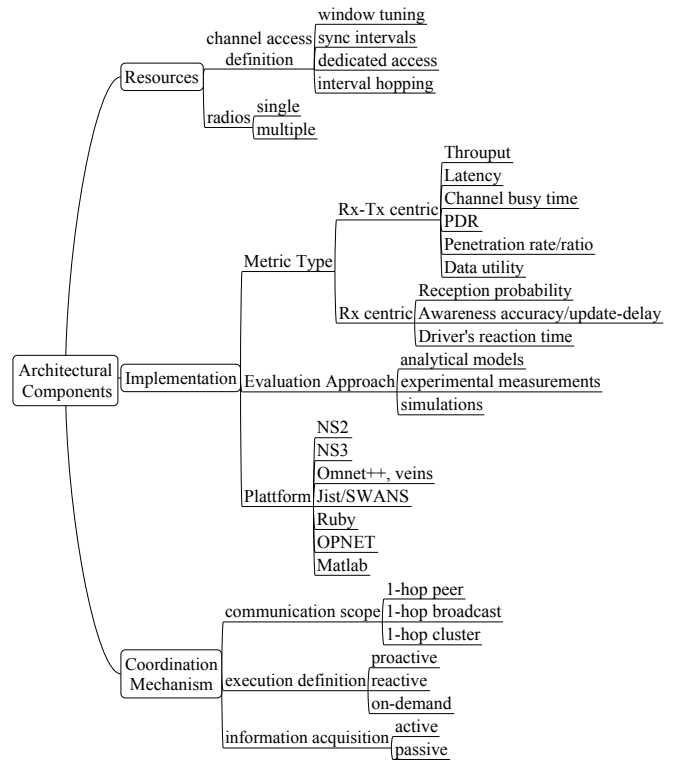

Fig. 5: Classification of Beaconing with respect to the Architectural Components including coordination mechanisms, implementation logic and available resources

The coordination mechanism characterizes the interaction of a vehicle with its neighbors. It is specified using three parameters: 1) communication scope, 2) execution definition and 3) information acquisition. Communication scope specifies the area for which beaconing is considered applicable. It implies that the 1-hop range can be further classified based on the region of interest. Execution definition specifies the time of action for the beaconing approach such as a) reactive, b) proactive, and c) on-demand. The information acquisition definition shows the type of information i.e. active or passive.

The underlying assumptions of the available architecture include number of radios and channel access definition. The beaconing standards allow for one radio for multi-channel access. However, in some cases, dual radio setup is also possible in order to increase the system capacity. Depending upon the number of radios, the control channels can be accessed in different ways. We represent this aspect by using channel access definition.

Implementation of beaconing at a system level can be characterized by studying the local implementation details. We classify implementation details by using four parameters: 1) evaluation approach, 2) metric type, and 3) platform.

### B. Comparison Among Beaconing Approaches

This section compares the beaconing approaches by analysing the strengths and weaknesses of different parameters as specified in Fig. 5. Table IV summarizes the comparison.

*1) Communication Scope:* Communication scope (CS) defines the applicability of beaconing to a logical organization of vehicles in a transmission range. It includes 1) 1-hop broadcast, which specifies all the vehicles in a transmission range. 2) 1-hop cluster, which is a dynamic organization consisting of group/s of vehicles. For instance, in [50], two distinct ranges are specified that require different transmit rate control. That is, vehicles within 100-meter range form a logical inner cluster with higher awareness requirement whereas, vehicles within 300-meter range form an outer cluster. 3) The 1-hop peer communication scope is restricted to unique vehicles within the 1-hop range. For instance, to identify a potential collision partner during mobility [53], and lane change applications [34] etc. The 1-hop cluster and 1-hop peer scope are suitable for specific safety situations. 4) Multi-hop CS has the ability to extend communication to the carrier sense range using multi-hop transmission [39]. The aim is to reduce network beaconing load while increasing transmission range without the need of increasing transmission power.

*2) Execution Definition:* Execution definition (ED) refers to the type of beaconing. It can take three values: a) reactive, b) proactive, and b) on-demand. Reactive defines beaconing after event detection. Reactive approaches can be classified as instances of event-driven mechanisms, in which events relate to a particular communication problem, application demand or safety condition. Reactive execution is easy to implement and produces deterministic results because the adaptive behavior is based on currently acquired information. Proactive execution is based on future estimates or models to prevent a communication problem or safety situation. For instance, saving the useful amount of CCH by preventing congestion and predicting a collision partner before a collision happens. However, the estimations and models are non-deterministic and involve certain assumptions, which might not reflect the actual state. Finally, on-demand execution defines the request mechanism for a particular adaptive behavior from neighbors [50]. Specifically, a high message frequency could be requested from a vehicle having highly variable longitudinal and lateral dynamics. On-demand beaconing execution has more implementation complexity with an additional request and response mechanism.

*3) Information Acquisition Definition:* Information acquisition definition (IAD) is the scope of information used for decision making. That is, it can be active or passive. Active IAD requires information sharing and has a wide scope indicative of neighbor vehicle information. It is based on a closed-loop control theory. Approaches with active IAD are capable of solving issues such as fairness [33], [3], [4], prioritization -[38] and preventing potential hazardous situations [50], [40]. The granularity of IAD varies as a) 1-hop, and b) multi-hop. Approaches with 1-hop IAD restricts the scope of acquired information to the 1-hop range. The 1-hop range specifies the standard range [9] or can be explicitly defined by the beaconing approach [50]. Multi-hop IAD [39] extends the information scope to the CS range. In the context of safety applications, multi-hop IAD affects information freshness that is critical for safety scenarios. Contrastingly, passive IAD depends on the local information. Its implementation is relatively simple, and it assumes a uniform view of the network. Therefore, it is not suitable for addressing dissemination problems such as fairness. However, it can be used effectively to address issues of general interest such as CCH load reduction and safety applications such as collision warning scenarios.

*4) Number of Radios:* Multi-channel operation for VANETS is designed for single and dual radio setups [13], [58]. Most of the approaches follow the standard single radio setup. Single radio setups are suitable for decentralized spectrum sharing. However, as described in previous sections, switching between CCH and SCH is time synchronized. It implies that during communication the available channel capacity decreases to half [14]. All approaches in this study use single radio except for BaaS [50]. Dual radio setup requires permanent tuning of one radio to the CCH while channel switching is allowed on the second radio. Use of dual radio increases the CCH capacity. Nonetheless, using dual radio for multi-channel access with no centralized scheduler introduces adjacent channel interference (ACI) [59], [60]. In addition, system complexity also increases with the need of dual radio management services. One of the countermeasures used in BaaS [50] against ACI is to reduce the transmission power in one of the radios, as a trade-off between reduced ACI and increased transmission range.

*5) Channel Access Definition:* Channel access definition (CAD) specifies the use of synchronization (sync) intervals and contention windows. There are different ways to access the CCH besides the synchronization interval defined in the standards [9], [13]. Most beaconing approaches use the default synch intervals. However, dedicated access [50], interval hopping [42], [43] and window tuning [45] are also proposed in literature. The main aim of different CAD is to improve neighbor awareness by minimizing collisions. Dedicated access replaces channel switching with a permanent CCH access to increase neighbor awareness of only one neighbor. Interval hopping [42] uses a heuristic of hoping to distinct firings interval to avoid synchronized collisions. However, it requires a steady state of vehicles for which firing intervals are evenly distributed in time space. Finally, window tuning [45], [46], [47] the increases the contention window for beaconing in order to provide time diversity in accessing a channel and hence minimizing collisions.

*6) Evaluation Approach:* Evaluation approaches (EA) are the mechanisms to evaluate performance of beaconing i.e. 1) analytical models, 2) simulation and 3) real implementation. Analytical models provide an abstract view of the system. It is a cost effective mechanism and provides a prompt evaluation. However, it does not represent dynamic properties of a real system. Analytical models lack randomness and heavily depend on the underlying assumptions. In order to validate an analytical model, the concepts need to be field-tested or simulated. For instance, in [47], the authors prove the suitability of higher contention windows during high-density networks. However, the extent of limits for contention window adaptation requires evaluation in a realistic network. Therefore, it is recommended to use analytical models alongside simulations and field-tests. On the other hand, simulations are used to prove the validity of a proposed analytical model

TABLE IV: Comparison of beaconing approaches with respect to the architectural characteristics

| Cat. | Ref | CS | ED | IAD | Radio/s | CAD | EA | Metric type |
|---|---|---|---|---|---|---|---|---|
| MFC | [27] | 1-hop broadcast | Reactive | Passive | Single (W2) | Synch intervals | Simulation | Rx centric |
| | [31] | 1-hop peer | Reactive | Passive | Single (W2) | Synch intervals | Simulation | Rx centric |
| | [32] | 1-hop broadcast | Reactive | Passive | Single (W2) | Synch intervals | Simulation | Rx-Tx centric |
| | [4] | 1-hop broadcast | Reactive | Act(1-hop) | Single (W2) | Synch intervals | Simulation | Rx centric |
| | [33] | 1-hop broadcast | Reactive | Act(multihop) | Single (W2) | Synch intervals | Simulation | Offered Load |
| | [34] | 1-hop peer | Proactive | Passive | Single (W2) | Synch intervals | Simulation | Rx centric |
| | [35] | 1-hop broadcast | Reactive | Passive | Single (W2) | Synch intervals | Simulation | Rx-Tx centric |
| | [37] | 1-hop broadcast | Reactive | Act(1-hop) | Single (W2) | Synch intervals | Simulation | Rx-Tx centric |
| TPC | [38] | 1-hop broadcast | Reactive | Act(multihop) | Single (W2) | Synch intervals | Simulation | Rx centric |
| | [40] | 1-hop broadcast | Reactive | Passive | Single (W2) | Synch intervals | Simulation | Rx centric |
| | [39] | Multi-hop | Reactive | Act(multihop) | Single (W2) | Synch intervals | Simulation | Rx-Tx centric |
| | [41] | 1-hop broadcast | Reactive | Act(1-hop) | Single (W2) | Synch intervals | Simulation | Rx-Tx centric |
| Misc | [42] | 1-hop broadcast | Reactive | Act(1-hop) | Single (W2) | Interval hopping | Simulation | Rx-Tx centric |
| | [43] | 1-hop broadcast | Reactive | Act(1-hop) | Single (W2) | Interval hopping | Simulation | Rx-Tx centric |
| | [44] | 1-hop broadcast | Reactive | Passive | Single (W2) | Synch intervals | Simulation | Rx centric |
| | [45] | 1-hop broadcast | Proactive | Passive | Single (W2) | Window tuning | Analy. model | Rx-Tx centric |
| | [46] | 1-hop broadcast | Reactive | Act(1-hop) | Single (W2) | Window tuning | Simulation | Rx-Tx centric |
| | [47] | 1-hop broadcast | Proactive | Passive | Single (W2) | Window tuning | analy./sim. | Rx-Tx centric |
| | [48] | 1-hop broadcast | Reactive | Passive | Single (W2) | Synch intervals | Simulation | Rx-Tx centric |
| | [49] | 1-hop broadcast | Pro/Reac | Act(1-hop) | Single (W2) | Synch intervals | Simulation | Rx-Tx centric |
| Hybrid | [51] | 1-hop broadcast | Reac/pro | Passive | Single (W2) | Synch intervals | Real | Rx centric |
| | [50] | 1-hop cluster | Reac/On-dem | Act(1-hop) | Dual (T2) | Dedicated Access | Simulation | Rx centric |
| | [53] | 1-hop peer | Reac /Reac | Passive | Single (W2) | Synch intervals | Simulation | Reliability |
| | [28] | 1-hop peer | Pro/Pro | Passive | Single (W2) | Synch intervals | Simulation | Rx centric |
| | [55] | 1-hop broadcast | Reactive | Passive | Single (W2) | Synch intervals | Simulation | Rx-Tx centric |

CS: communication scope, ED: execution definition, IAD:information acquisition definition, CAD: channel access definition, EA: evaluation approach, Act: active, pro: proactive, reac: reactive, on-dem: on-demand, Synch: synchronization, Analy: analytical, Rx: receiver, Rx-Tx: receiver-transmitter

[45]. Simulations provide randomness in the environment and produce accurate results with less abstraction of underlying assumptions. However, too many details may lead to an unmanageable simulation environment. Finally, evaluations using test-beds are rare in VANETs [52]. Such evaluation is more costly and has the minimal assumption for a distinct test case. However, for large systems, designing a test-bed is not feasible and it must be accompanied by simulation results to make evaluation more dynamic.

*7) Metric Type:* Metric type classifies metrics to indicate the performance gains of a proposed beaconing approach. That is, a traditional receiver-transmitter (Rx-Tx) centric metric is used to show the overall system performance such as throughput, packet delivery ratio, and latency etc. On the other hand, receiver (Rx) centric metrics for beaconing are used to evaluate the application-level performance of beaconing approaches. For the evaluation of safety applications, Rx centric metrics are more desirable, which can capture specific effects such as driver's reaction time and update delays to name but a few. In Fig. 5, some example metrics under each category are shown. Note that, penetration rate/ratio and data utility are the metrics used for non-safety TIS. The former specifies the speed and coverage of TIS data while the later indicates the benefit of selected data for other vehicles upon reception.

## VIII. OPEN CHALLENGES

We observe that certain aspects in adaptive beaconing can be improved as discussed in the following.

### A. Beaconing Approaches for ITS Co-existence

The literature about cooperative communication emphasizes on the data that represents public interest [61]. However, routing protocols representing individual interest [62] also require accurate neighbor position for applications such as infotainment [63]. However, the requirements of position updates are different for safety applications and for applications that use routing protocols. As aforementioned, safety applications aim for accurate position updates of immediate neighbors. On the contrary, unicast routing protocols require a higher degree of awareness of distant vehicles to reduce the number of hops in a unicast path. In mobile ad-hoc networks (MANETs), it has been shown that adaptive beaconing can be effective in improving topology awareness as well as unicast path accuracy [64]. However, considering contradictory application requirements in VANETs, it remains a challenge to design beaconing approaches for co-exiting ITS applications.

### B. Flexible Channelization for Spatial Reuse

According to [65], an assigned wireless spectrum can be underutilized if the communication activity is sporadic. Likewise, 5 GHz spectrum for VANETs can be considered as a special case of an underutilized spectrum having the majority of communication activity in the CCH. That is, subject to empirical evaluation, multiple SCHs are underutilized due to 1) deterministic nature of traffic transmission and 2) flexibility of switching to any SCH upon availability. However, 802.11p multi-channel access is inflexible in the choice of channel selection for periodic beacons. Considering scarce CCH resources and a high-priority transmission on the CCH, there is a need for more flexible approaches that allows opportunistic access to the SCH for CCH data. However, such opportunistic access requires a cognitive behavior [66] for channel switching to assess a suitable channel and more importantly a seamless mechanism for decentralized channel switching.

### C. Heterogeneous Network Communication for Scalability

Most beaconing approaches address the problem of scalability under scarce CCH resources and increase in vehicle density. Alternatively, in situations of sparse network connectivity, these approaches will not be suitable because they assume a certain level of vehicular density. As a possible solution for improved connectivity, frequencies assigned for analog television could be explored. As of now, these frequencies have started to free up due to the shift towards digital television services. These unused frequencies are called white-spaces and can be used for long range ITS communications. In addition, use of satellite communication has been proposed for non-safety applications under sparse connectivity [67]. However, a critical implication of satellite network communication is the propagation delay that may affect the information freshness. Further investigation is required in heterogeneous communication patterns that can reduce the effects of sparse connectivity and retain information freshness at the same time.

### IX. CONCLUSION

Intelligent transportation systems have strict beacon dissemination requirements, which are typically addressed through various beaconing approaches. This paper has contributed by conducting a survey of the adaptive beaconing approaches designed for effective ITS. Specifically, it first identified the performance requirements of vehicular applications. Then, the information used as input by the beaconing approaches and the subsequent choice of a control mechanism is explored to propose a design-based beaconing taxonomy. The salient features of a number of beaconing approaches under each category of MFC, TPC, Miscellaneous and Hybrid are surveyed and key observations about each category are listed. We further evaluated the capabilities of beaconing approaches on a qualitative scale by considering parameters including beaconing load, congestion control strategy, fairness, reliability, data utility distribution and co-existing message dissemination evaluation. In addition, the variations in the architectural and implementation aspects among the beaconing approaches are used to highlight similarities and variations among them. Along the lines of these contributions, we presented open challenges to conclude the paper.

### X. ACKNOWLEDGMENT

This work is supported by High Impact Research, University of Malaya, and MoHE (UM.C/HIR/MOHE/FCSIT/09).

### REFERENCES

[1] C. D. Wang and J. P. Thompson, "Apparatus and method for motion detection and tracking of objects in a region for collision avoidance utilizing a real-time adaptive probabilistic neural network," Mar. 18 1997, uS Patent 5,613,039.

[2] C. V. S. C. Consortium *et al.*, "Vehicle safety communications project: task 3 final report: identify intelligent vehicle safety applications enabled by dsrc," *National Highway Traffic Safety Administration, US Department of Transportation, Washington DC*, 2005.

[3] J. He, H.-H. Chen, T. M. Chen, and W. Cheng, "Adaptive congestion control for dsrc vehicle networks," *IEEE communications letters*, vol. 14, no. 2, pp. 127–129, 2010.

[4] W. Guan, J. He, L. Bai, and Z. Tang, "Adaptive congestion control of dsrc vehicle networks for collaborative road safety applications," in *Local Computer Networks (LCN), 2011 IEEE 36th Conference on*. IEEE, 2011, pp. 913–917.

[5] M. R. Jabbarpour, R. M. Noor, R. H. Khokhar, and C.-H. Ke, "Cross-layer congestion control model for urban vehicular environments," *Journal of Network and Computer Applications*, vol. 44, pp. 1–16, 2014.

[6] M. Sepulcre, J. Mittag, P. Santi, H. Hartenstein, and J. Gozalvez, "Congestion and awareness control in cooperative vehicular systems," *Proceedings of the IEEE*, vol. 99, no. 7, pp. 1260–1279, 2011.

[7] T. L. Willke, P. Tientrakool, and N. F. Maxemchuk, "A survey of inter-vehicle communication protocols and their applications," *Communications Surveys & Tutorials, IEEE*, vol. 11, no. 2, pp. 3–20, 2009.

[8] K. Z. Ghafoor, J. Lloret, K. A. Bakar, A. S. Sadiq, and S. A. B. Mussa, "Beaconing approaches in vehicular ad hoc networks: a survey," *Wireless personal communications*, vol. 73, no. 3, pp. 885–912, 2013.

[9] "Ieee standard for information technology– local and metropolitan area networks– specific requirements– part 11: Wireless lan medium access control (mac) and physical layer (phy) specifications amendment 6: Wireless access in vehicular environments," *IEEE Std 802.11p-2010 (Amendment to IEEE Std 802.11-2007 as amended by IEEE Std 802.11k-2008, IEEE Std 802.11r-2008, IEEE Std 802.11y-2008, IEEE Std 802.11n-2009, and IEEE Std 802.11w-2009)*, pp. 1–51, July 2010.

[10] E. ETSI, "302 665, 3 (2010) intelligent transport systems (its)," *Communications Architecture*.

[11] R. Uzcategui and G. Acosta-Marum, "Wave: a tutorial," *Communications Magazine, IEEE*, vol. 47, no. 5, pp. 126–133, 2009.

[12] "Ieee standard for wireless access in vehicular environments (wave) - networking services," *IEEE Std 1609.3-2010 (Revision of IEEE Std 1609.3-2007)*, pp. 1–144, Dec 2010.

[13] "Ieee standard for wireless access in vehicular environments (wave)– multi-channel operation," *IEEE Std 1609.4-2010 (Revision of IEEE Std 1609.4-2006)*, pp. 1–89, Feb 2011.

[14] C. Campolo and A. Molinaro, "Multichannel communications in vehicular ad hoc networks: a survey," *Communications Magazine, IEEE*, vol. 51, no. 5, pp. 158–169, 2013.

[15] "Ieee standard for information technology–telecommunications and information exchange between systems local and metropolitan area networks–specific requirements part 11: Wireless lan medium access control (mac) and physical layer (phy) specifications," *IEEE Std 802.11-2012 (Revision of IEEE Std 802.11-2007)*, pp. 1–2793, March 2012.

[16] D. Jiang, Q. Chen, and L. Delgrossi, "Optimal data rate selection for vehicle safety communications," in *Proceedings of the fifth ACM international workshop on VehiculAr Inter-NETworking*. ACM, 2008, pp. 30–38.

[17] T. ETSI, "Intelligent transport systems (its); vehicular communications; basic set of applications; definitions," Tech. Rep. ETSI TR 102 638, Tech. Rep., 2009.

[18] J. L. Gabbard, G. M. Fitch, and H. Kim, "Behind the glass: Driver challenges and opportunities for automotive applications," *Proceedings of the IEEE*, vol. 102, no. 2, pp. 124–136, 2014.

[19] M. Sepulcre, J. Gozalvez, and J. Hernandez, "Cooperative vehicle-to-vehicle active safety testing under challenging conditions," *Transportation research part C: emerging technologies*, vol. 26, pp. 233–255, 2013.

[20] S. Céspedes, N. Lu, and X. Shen, "Vip-wave: On the feasibility of ip communications in 802.11 p vehicular networks," *Intelligent Transportation Systems, IEEE Transactions on*, vol. 14, no. 1, pp. 82–97, 2013.

[21] Y. Liu, J. Niu, J. Ma, and W. Wang, "File downloading oriented roadside units deployment for vehicular networks," *Journal of Systems Architecture*, vol. 59, no. 10, pp. 938–946, 2013.

[22] K. Ota, M. Dong, S. Chang, and H. Zhu, "Mmcd: cooperative downloading for highway vanets," *Emerging Topics in Computing, IEEE Transactions on*, vol. 3, no. 1, pp. 34–43, 2015.

[23] ——, "Mmcd: Max-throughput and min-delay cooperative downloading for drive-thru internet systems," in *Communications (ICC), 2014 IEEE International Conference on*. IEEE, 2014, pp. 83–87.

[24] H. Zhang, Y. Ma, D. Yuan, and H.-H. Chen, "Quality-of-service driven power and sub-carrier allocation policy for vehicular communication networks," *Selected Areas in Communications, IEEE Journal on*, vol. 29, no. 1, pp. 197–206, 2011.

[25] K. Yang, S. Ou, H. Chen, and J. He, "A multihop peer-communication protocol with fairness guarantee for ieee 802.16-based vehicular networks," *Vehicular Technology, IEEE Transactions on*, vol. 56, no. 6, pp. 3358–3370, 2007.

[26] F. Schmidt-Eisenlohr, M. Torrent-Moreno, J. Mittag, and H. Hartenstein, "Simulation platform for inter-vehicle communications and analysis of periodic information exchange," in *Wireless on Demand Network Systems and Services, 2007. WONS'07. Fourth Annual Conference on*. IEEE, 2007, pp. 50–58.

[27] R. Schmidt, T. Leinmuller, E. Schoch, F. Kargl, and G. Schafer, "Exploration of adaptive beaconing for efficient intervehicle safety communication," *Network, IEEE*, vol. 24, no. 1, pp. 14–19, 2010.

[28] J. Gozalvez and M. Sepulcre, "Opportunistic technique for efficient wireless vehicular communications," *Vehicular Technology Magazine, IEEE*, vol. 2, no. 4, pp. 33–39, 2007.

[29] S. Narayanaswamy, V. Kawadia, R. S. Sreenivas, and P. Kumar, "Power control in ad-hoc networks: Theory, architecture, algorithm and implementation of the compow protocol," in *European wireless conference*, vol. 2002. Florence, Italy, 2002.

[30] P. Gupta and P. R. Kumar, "The capacity of wireless networks," *Information Theory, IEEE Transactions on*, vol. 46, no. 2, pp. 388–404, 2000.

[31] A. Boukerche, C. Rezende, and R. W. Pazzi, "Improving neighbor localization in vehicular ad hoc networks to avoid overhead from periodic messages," in *Global Telecommunications Conference, 2009. GLOBECOM 2009. IEEE*. IEEE, 2009, pp. 1–6.

[32] S. Rezaei, R. Sengupta, and H. Krishnan, "Reducing the communication required by dsrc-based vehicle safety systems," in *Intelligent Transportation Systems Conference, 2007. ITSC 2007. IEEE*. IEEE, 2007, pp. 361–366.

[33] T. Tielert, D. Jiang, Q. Chen, L. Delgrossi, and H. Hartenstein, "Design methodology and evaluation of rate adaptation based congestion control for vehicle safety communications," in *Vehicular Networking Conference (VNC), 2011 IEEE*. IEEE, 2011, pp. 116–123.

[34] A. Bohm, M. Jonsson, and E. Uhlemann, "Adaptive cooperative awareness messaging for enhanced overtaking assistance on rural roads," in *Vehicular Technology Conference (VTC Fall), 2011 IEEE*. IEEE, 2011, pp. 1–5.

[35] C. Sommer, O. K. Tonguz, and F. Dressler, "Traffic information systems: efficient message dissemination via adaptive beaconing," *Communications Magazine, IEEE*, vol. 49, no. 5, pp. 173–179, 2011.

[36] K. Ota, M. Dong, H. Zhu, S. Chang, and X. Shen, "Traffic information prediction in urban vehicular networks: A correlation based approach," in *Wireless Communications and Networking Conference (WCNC), 2011 IEEE*. IEEE, 2011, pp. 1021–1025.

[37] R. S. Schwartz, A. E. Ohazulike, C. Sommer, H. Scholten, F. Dressler, and P. Havinga, "On the applicability of fair and adaptive data dissemination in traffic information systems," *Ad hoc networks*, vol. 13, pp. 428–443, 2014.

[38] M. Torrent-Moreno, J. Mittag, P. Santi, and H. Hartenstein, "Vehicle-to-vehicle communication: fair transmit power control for safety-critical information," *Vehicular Technology, IEEE Transactions on*, vol. 58, no. 7, pp. 3684–3703, 2009.

[39] J. Mittag, F. Thomas, J. Härri, and H. Hartenstein, "A comparison of single-and multi-hop beaconing in vanets," in *Proceedings of the sixth ACM international workshop on VehiculAr InterNETworking*. ACM, 2009, pp. 69–78.

[40] B. Kloiber, J. Härri, T. Strang, F. Hrizi, C. Bonnet, C. Rico-Garcia, P. Matzakos, D. Krajzewicz, L. Bieker, and R. Blokpoel, "Dice the tx power-improving awareness quality in vanets by random transmit power selection," in *VNC*, 2012, pp. 56–63.

[41] X. Guan, R. Sengupta, H. Krishnan, and F. Bai, "A feedback-based power control algorithm design for vanet," in *2007 Mobile Networking for Vehicular Environments*. IEEE, 2007, pp. 67–72.

[42] T. Settawatcharawanit, S. Choochaisri, C. Intanagonwiwat, and K. Rojviboonchai, "V-desync: Desynchronization for beacon broadcasting on vehicular networks," in *Vehicular Technology Conference (VTC Spring), 2012 IEEE 75th*. IEEE, 2012, pp. 1–5.

[43] Y. Park and H. Kim, "Collision control of periodic safety messages with strict messaging frequency requirements," *Vehicular Technology, IEEE Transactions on*, vol. 62, no. 2, pp. 843–852, 2013.

[44] P. Shankar, T. Nadeem, J. Rosca, and L. Iftode, "Cars: Context-aware rate selection for vehicular networks," in *Network Protocols, 2008. ICNP 2008. IEEE International Conference on*. IEEE, 2008, pp. 1–12.

[45] C. Campolo, A. Molinaro, A. Vinel, and Y. Zhang, "Modeling prioritized broadcasting in multichannel vehicular networks," *Vehicular Technology, IEEE Transactions on*, vol. 61, no. 2, pp. 687–701, 2012.

[46] M. Di Felice, A. J. Ghandour, H. Artail, and L. Bononi, "On the impact of multi-channel technology on safety-message delivery in ieee 802.11 p/1609.4 vehicular networks," in *Computer Communications and Networks (ICCCN), 2012 21st International Conference on*. IEEE, 2012, pp. 1–8.

[47] R. Stanica, E. Chaput, and A.-L. Beylot, "Reverse back-off mechanism for safety vehicular ad hoc networks," *Ad Hoc Networks*, vol. 16, pp. 210–224, 2014.

[48] R. K. Schmidt, A. Brakemeier, T. Leinmüller, F. Kargl, and G. Schäfer, "Advanced carrier sensing to resolve local channel congestion," in *Proceedings of the Eighth ACM international workshop on Vehicular inter-networking*. ACM, 2011, pp. 11–20.

[49] J. Yoo, "Receiver-centric physical carrier sensing for vehicular ad hoc networks," *Multimedia Tools and Applications*, pp. 1–11, 2013.

[50] R. Lasowski and C. Linnhoff-Popien, "Beaconing as a service: a novel service-oriented beaconing strategy for vehicular ad hoc networks," *Communications Magazine, IEEE*, vol. 50, no. 10, pp. 98–105, 2012.

[51] C.-L. Huang, Y. P. Fallah, R. Sengupta, and H. Krishnan, "Adaptive intervehicle communication control for cooperative safety systems," *Network, IEEE*, vol. 24, no. 1, pp. 6–13, 2010.

[52] C.-L. Huang, R. Sengupta, H. Krishnan, and Y. P. Fallah, "Implementation and evaluation of scalable vehicle-to-vehicle safety communication control," *Communications Magazine, IEEE*, vol. 49, no. 11, pp. 134–141, 2011.

[53] M. Sepulcre, J. Gozalvez, J. Harri, and H. Hartenstein, "Application-based congestion control policy for the communication channel in vanets," *Communications Letters, IEEE*, vol. 14, no. 10, pp. 951–953, 2010.

[54] C.-L. Huang, Y. P. Fallah, R. Sengupta, and H. Krishnan, "Information dissemination control for cooperative active safety applications in vehicular ad-hoc networks," in *Global Telecommunications Conference, 2009. GLOBECOM 2009. IEEE*. IEEE, 2009, pp. 1–6.

[55] D. B. Rawat, D. C. Popescu, G. Yan, and S. Olariu, "Enhancing vanet performance by joint adaptation of transmission power and contention window size," *Parallel and Distributed Systems, IEEE Transactions on*, vol. 22, no. 9, pp. 1528–1535, 2011.

[56] R. S. Schwartz, A. E. Ohazulike, C. Sommer, H. Scholten, F. Dressler, and P. Havinga, "Fair and adaptive data dissemination for traffic information systems," in *Vehicular Networking Conference (VNC), 2012 IEEE*. IEEE, 2012, pp. 1–8.

[57] A. Böhm, M. Jonsson, and E. Uhlemann, "Co-existing periodic beaconing and hazard warnings in ieee 802.11 p-based platooning applications," in *Proceeding of the tenth ACM international workshop on Vehicular inter-networking, systems, and applications*. ACM, 2013, pp. 99–102.

[58] T. ETSI, "Its; harmonised channel specifications for intelligent transport systems operating in the 5 ghz range," *DTS-0040016*, 2009.

[59] D. Hu and S. Mao, "On co-channel and adjacent channel interference mitigation in cognitive radio networks," *Ad Hoc Networks*, vol. 11, no. 5, pp. 1629–1640, 2013.

[60] M. A. Qureshi, R. M. Noor, A. Shamim, S. Shamshirband, and K.-K. R. Choo, "A lightweight radio propagation model for vehicular communication in road tunnels," *PloS one*, vol. 11, no. 3, p. e0152727, 2016.

[61] M. Gerla, C. Wu, G. Pau, and X. Zhu, "Content distribution in vanets," *Vehicular Communications*, vol. 1, no. 1, pp. 3–12, 2014.

[62] S. A. A. SHAH, M. SHIRAZ, M. K. NASIR, and R. B. M. NOOR, "Unicast routing protocols for urban vehicular networks: review, taxonomy, and open research issues," *Journal of Zhejiang University Science C (Computers & Electronics)*, 2014. [Online]. Available: http://www.zju.edu.cn/jzus/iparticle.php?doi=10.1631/jzus.C1300332

[63] M. Oche, R. M. Noor, and J. I. Aghinya, "Network centric qos performance evaluation of iptv transmission quality over vanets," *Computer Communications*, vol. 61, pp. 34–47, 2015.

[64] Q. Chen, S. Kanhere, and M. Hassan, "Adaptive position update for geographic routing in mobile ad hoc networks," *Mobile Computing, IEEE Transactions on*, vol. 12, no. 3, pp. 489–501, March 2013.

[65] J. Marinho and E. Monteiro, "Cognitive radio: survey on communication protocols, spectrum decision issues, and future research directions," *Wireless Networks*, vol. 18, no. 2, pp. 147–164, 2012. [Online]. Available: http://dx.doi.org/10.1007/s11276-011-0392-1

[66] E. Ahmed, M. Shiraz, and A. Gani, "Spectrum-aware distributed channel assignment for cognitive radio wireless mesh networks," *Malaysian Journal of Computer Science*, vol. 26, no. 3, pp. 232–250, 2013.

[67] B. Kloiber, T. Strang, H. Spijker, and G. Heijenk, "Improving information dissemination in sparse vehicular networks by adding satellite communication," in *Intelligent Vehicles Symposium (IV), 2012 IEEE*. IEEE, 2012, pp. 611–617.